# Comparison of parallel sorting algorithms

Darko Božidar and Tomaž Dobravec

*Faculty of Computer and Information Science, University of Ljubljana, Slovenia*

Technical report

**TABLE OF CONTENTS**

# 1. INTRODUCTION

In our study we implemented and compared seven sequential and parallel sorting algorithms: bitonic sort, multistep bitonic sort, adaptive bitonic sort, merge sort, quicksort, radix sort and sample sort. Sequential algorithms were implemented on a central processing unit using C++, whereas parallel algorithms were implemented on a graphics processing unit using CUDA platform. We chose these algorithms because to the best of our knowledge their sequential and parallel implementations were not yet compared all together in the same execution environment. We improved the above mentioned implementations and adopted them to be able to sort input sequences of arbitrary length. We compared algorithms on six different input distributions, which consisted of 32-bit numbers, 32-bit key-value pairs, 64-bit numbers and 64-bit key-value pairs.

In this report we give a short description of seven sorting algorithms and all the results obtained by our tests.

# 2. THE ALGORITHMS

In this chapter a short description of each sorting algorithm is presented. For detailed information about algorithms see the cited references.

## 2.1 BITONIC SORT

Bitonic sort was designed by Ken E. Batcher [3] and because of its simplicity is one of the most studied algorithms on GPU. It falls into the group of *sorting networks*, which means, that the sequence and direction of comparisons is determined in advance and is independent of input sequence. Bitonic sort is based on a bitonic sequence [3, 4, 5].

## 2.2 MULTISTEP BITONIC SORT

Parallel implementation of bitonic sort is very efficient when sorting short sequences, but it becomes slower when sorting very long sequences, because shared memory size is limited and long bionic sequences cannot be saved into it. In order to merge long bionic sequences, global memory has to be used instead of shared memory. Furthermore, global memory has to be accessed for every step of bionic merge. In order to increase the speed of sort, multiple steps of bitonic merge have to be executed with a single kernel invocation. This can be achieved with *multistep bitonic sort* [4].

## 2.3 IBR BITONIC SORT

*Interval Based Rearrangement (IBR) bitonic sort* [5] is based on *adaptive bitonic sort* implemented by Bilardi et. al. [20]. Adaptive bitonic sort operates on the idea, that every bitonic sequence can be merged, if first half of a bitonic sequence and second half of a bitonic sequence are ring-shifted by a value called *q*. Value *q* can be found with a variation of a binary search. In order to exchange elements in a bitonic merge step in time O(log n), a variation of a binary tree (*bitonic tree* [5, 20]) is used.

## 2.4 MERGE SORT

Merge sort is based on the *divide-and-conquer* approach. It follows this approach by splitting the sequence into multiple subsequences, sorting them and then merging them into sorted sequence. If the algorithm for merging sorted sequences is stable, than the whole merge sort algorithm is also stable [22]. In our tests we used a variation of a parallel merge sort presented by Satish in [2].

## 2.5 QUICKSORT

Similarly as merge sort, quicksort is also based on the *divide-and-conquer* basis. In *divide* step, the *pivot* has to be determined. This pivot is used to partition the input sequence A[p...r] into 2 subsequences. When the partitioning is done, pivot has to be placed in A[q], where all the elements in subsequence A[p...q − 1] have to be lower or equal to A[q] and all the elements of A[q + 1...r] have to be greater than A[q]. In step *conquer,* subsequences A[p...q-1] and A[q + 1...r] have to be recursively sorted as described above, until subsequences of length 1 are obtained. This recursive procedure sorts the entire sequence T [p...r] [22, 23].

In out tests we used the parallel quicksort designed by Cederman et. al. [1]. This algorithm first determines an initial pivot, which is calculated as an average of a minimum and a maximum element of an input sequence, as suggested by Singleton et. al. [24]. Minimum and maximum values were calculated using parallel reduction kernel by Harris [7], which improved the performance of sort [1]. Performance was further improved by Sengupta's et al. scan [8]. We also introduced a novel approach for determining a pivot.

## 2.6 RADIX SORT

Radix sort is one of the fastest sorting algorithms for short keys and is the only sorting algorithm in this report which is not comparison based. Its sequential variation first splits the elements being sorted (numbers, words, dates, ...) into *d r*-bit digits. The elements are then sorted from least to most significant digit. For this task, the sorting algorithm has to be stable, in order to preserve the order of elements with duplicate digits. For a good performance an effective sorting algorithm has to be used, which is usually *counting sort* [2, 19, 22]. We improved the performance of a parallel algorithm by using Harris's et al. binary scan [9].

## 2.7 SAMPLE SORT

Sample sort is a sorting algorithm, which splits the input sequence into multiple smaller buckets, until they are small enough to be sorted by alternative sort. For the parallel implementation we chose variation of the sample sort by Dehne et. al. [6], because it is more robust for sorting different types of input sequences than Leischner's et. al. [13, 19].

# 3. TESTING ENVIRONMENT

We compared the efficiency of sorting algorithms on the CPU *Intel Core i7-3770K* with a frequency of 3.5GHz. For parallel computing we used the GPU *GeForce GTX670* with 2GB of memory. Algorithms were tested on 6 input distributions sorting 32-bit keys, 32-bit key-value pairs, 64-bit keys and 64-bit key-value pairs. For random number generator we chose *Mersenne Twister* [25]. Input sequences were of length from $2^{15}$ to $2^{25}$ for 32-bit numbers and from $2^{15}$ to $2^{24}$ for 64-bit numbers. To test the efficiency of algorithms on non-regular sequences (i.e., the sequences with the length not equal to the power of 2), our test sets also included sequences of length $2^n + 2^{n-1}$ (for n = 15, . . . , 24). We represented the speed of the sort with so called *sort rate*, which is the number of sorted elements (in millions) per second (M/s). For a given sequence we ran each algorithm for 50-times and calculated the average value of the time required. We timed only the execution of the sort without memory transfer to and from the device. This is due to the fact, that sorting algorithm is usually just a part of more complex algorithm, which means that the data is already located on the GPU, so this kind of measuring became a standard in the literature [1].

# 4. RESULTS

Images below contain the results of sequential and parallel sorts. All sorting algorithms were tested on six different input distributions (uniform, Gaussian, zero, bucket, sorted, sorted descending), which consisted of 32-bit numbers, 32-bit key-value pairs, 64-bit numbers and 64-bit key-value pairs. X axis contains the binary logarithm of sequence length, while Y axis contains sort rate, which is the number of sorted elements (in millions) per second (M/s).
Results for sequential sorts of zero input distribution don't contain quicksort, because it achieves the speed of only 0.3 M/s. The same goes for results of parallel sort of zero input distribution, where quicksort achieves the speed of 35.000 M/s, because it only needs to find minimum and maximum values.

**Speedup**

Speedup = the quotient between the speed of the parallel algorithm and the speed of the corresponding sequential algorithm.

X-axis: Binary logarithm of the sequence length, Y-axis: speedup.

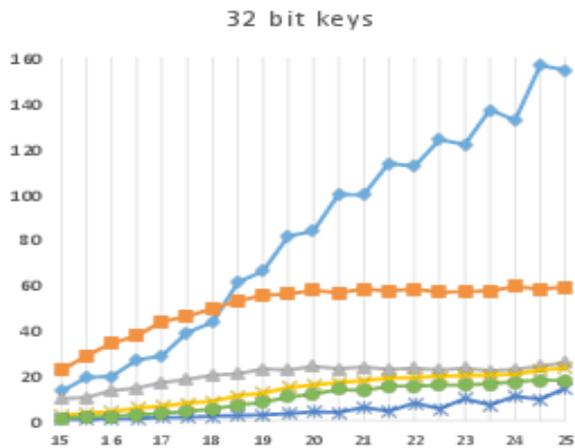
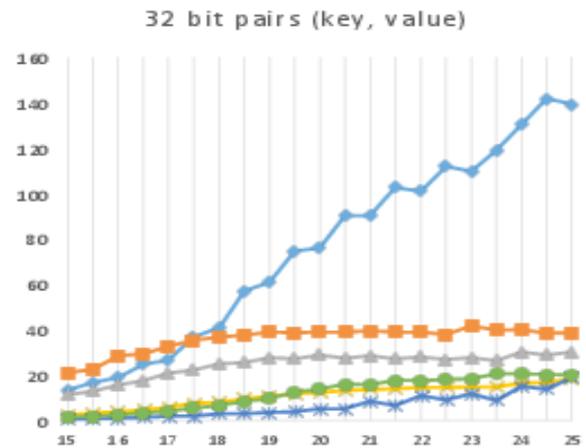
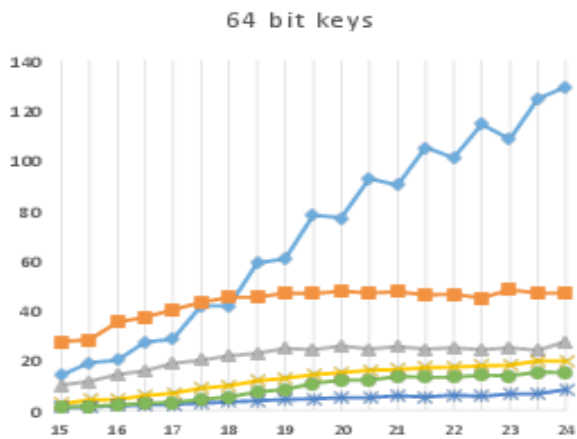
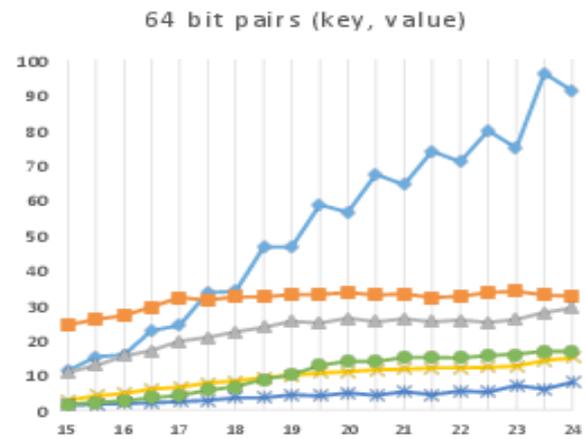

## Parallel, 32 bit, keys only
The sort rate of parallel algorithms when sorting sequences of 32-bit keys.
X-axis: Binary logarithm of the sequence length, Y-axis: sort rate in M/s.

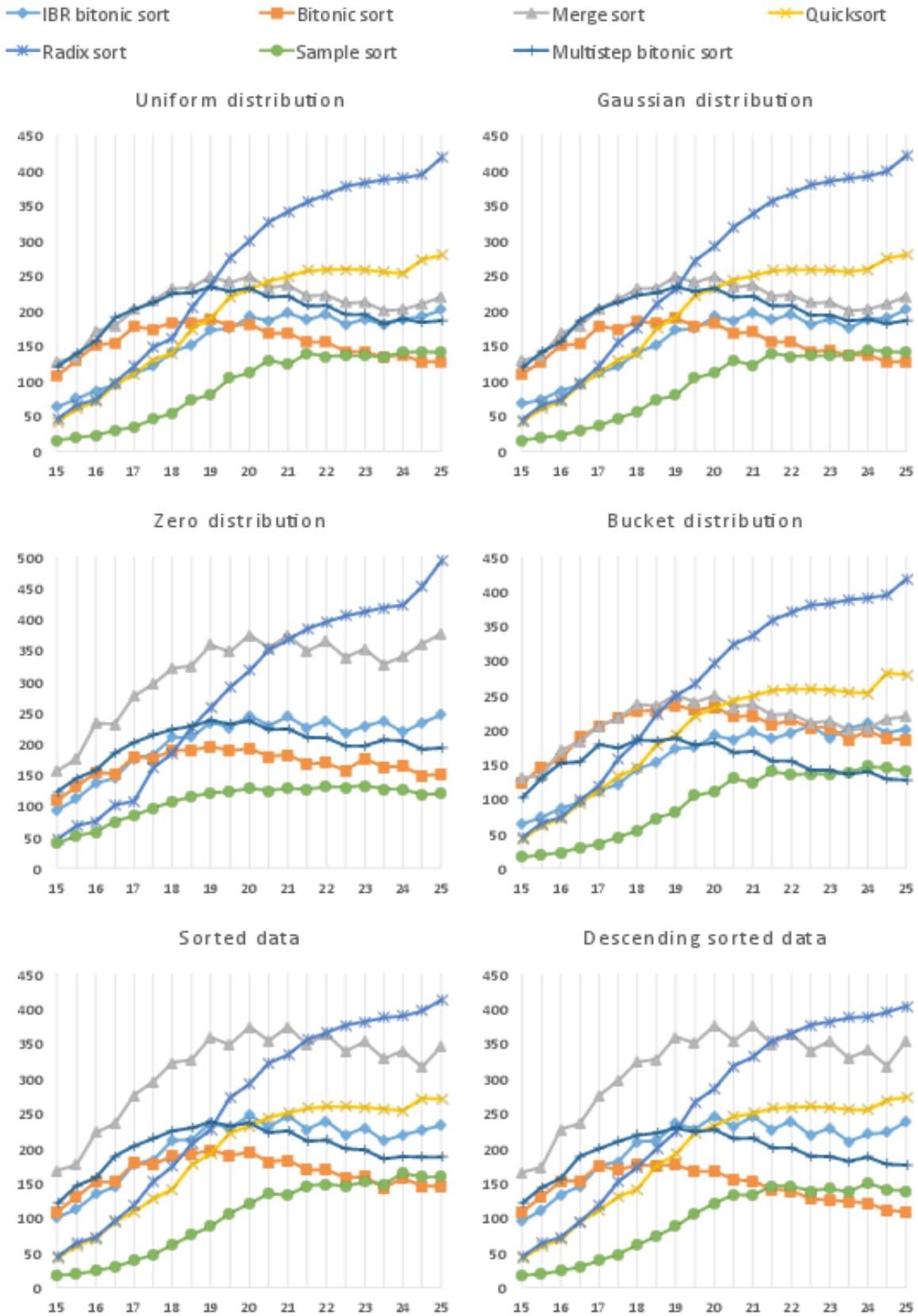

## Parallel, 32 bit, (key, value) pairs
The sort rate of parallel algorithms when sorting sequences of 32-bit (key, value) pairs.
X-axis: Binary logarithm of the sequence length, Y-axis: sort rate in M/s.

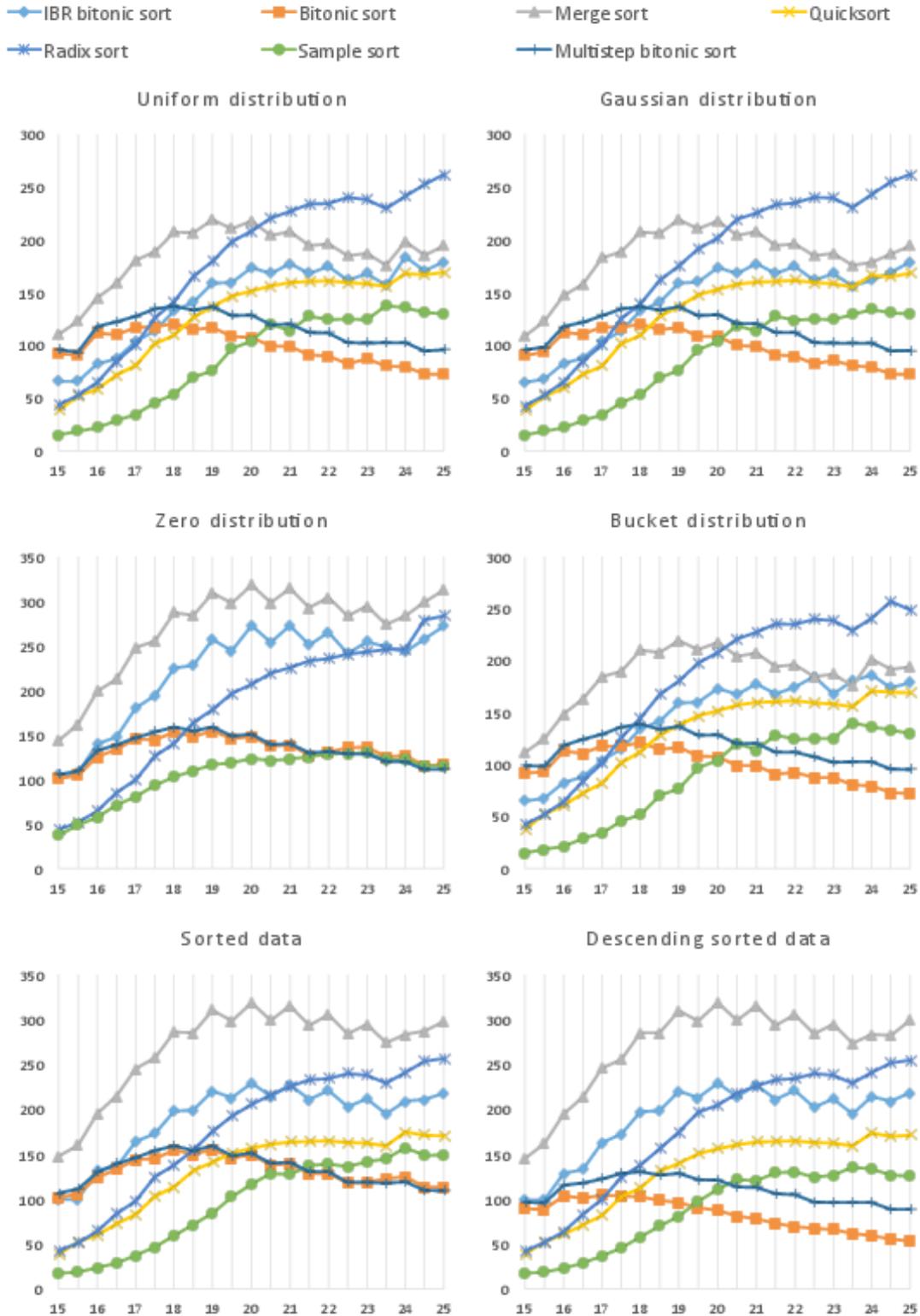

## Parallel, 64 bit, keys only
The sort rate of parallel algorithms when sorting sequences of 64-bit keys.
X-axis: Binary logarithm of the sequence length, Y-axis: sort rate in M/s.

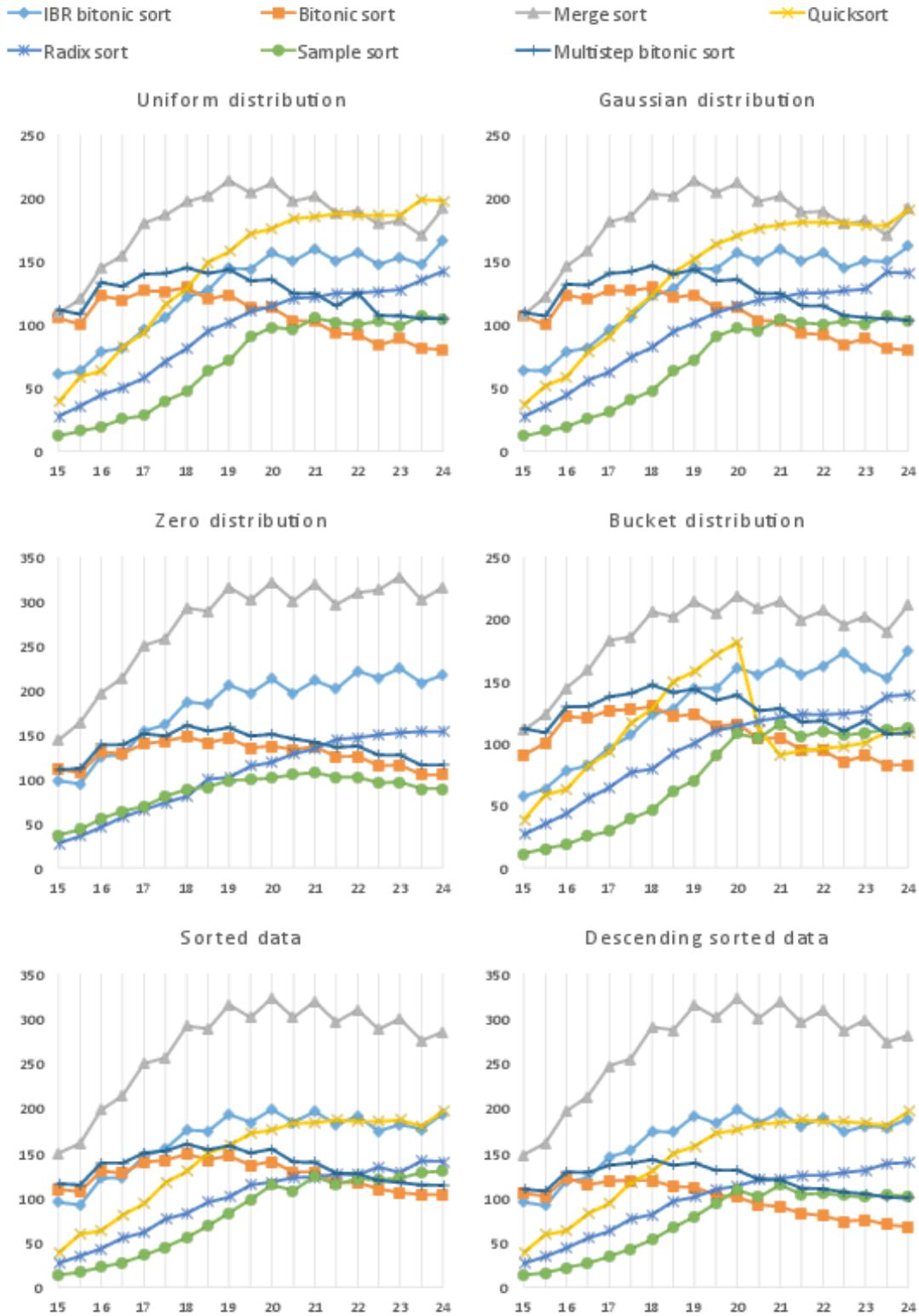

## Parallel, 64 bit, (key, value) pairs
The sort rate of parallel algorithms when sorting sequences of 64-bit (key, value) pairs.
X-axis: Binary logarithm of the sequence length, Y-axis: sort rate in M/s.

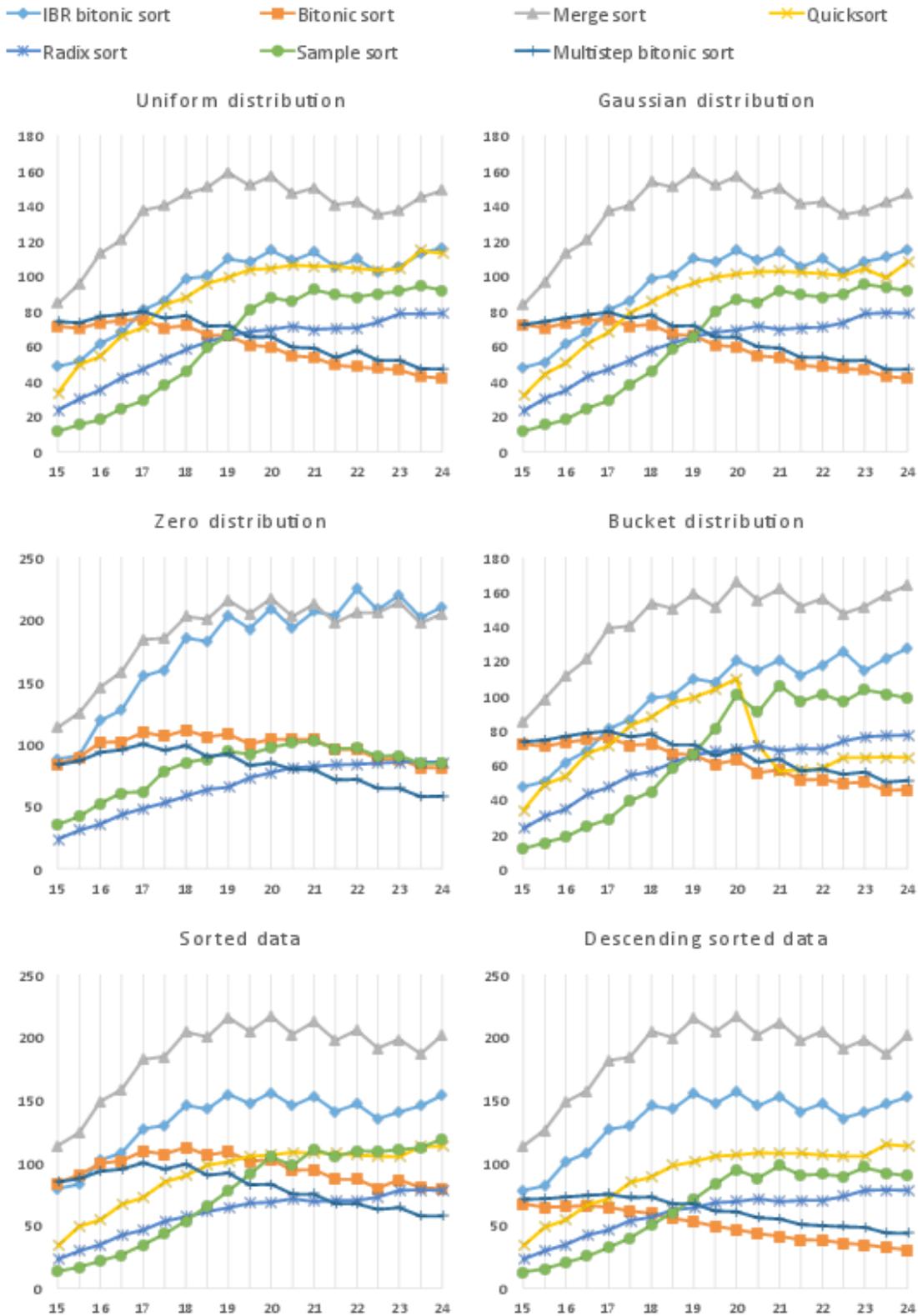

## Sequential, 32 bit, keys only
The sort rate of sequential algorithms when sorting sequences of 32-bit keys.
X-axis: Binary logarithm of the sequence length, Y-axis: sort rate in M/s.

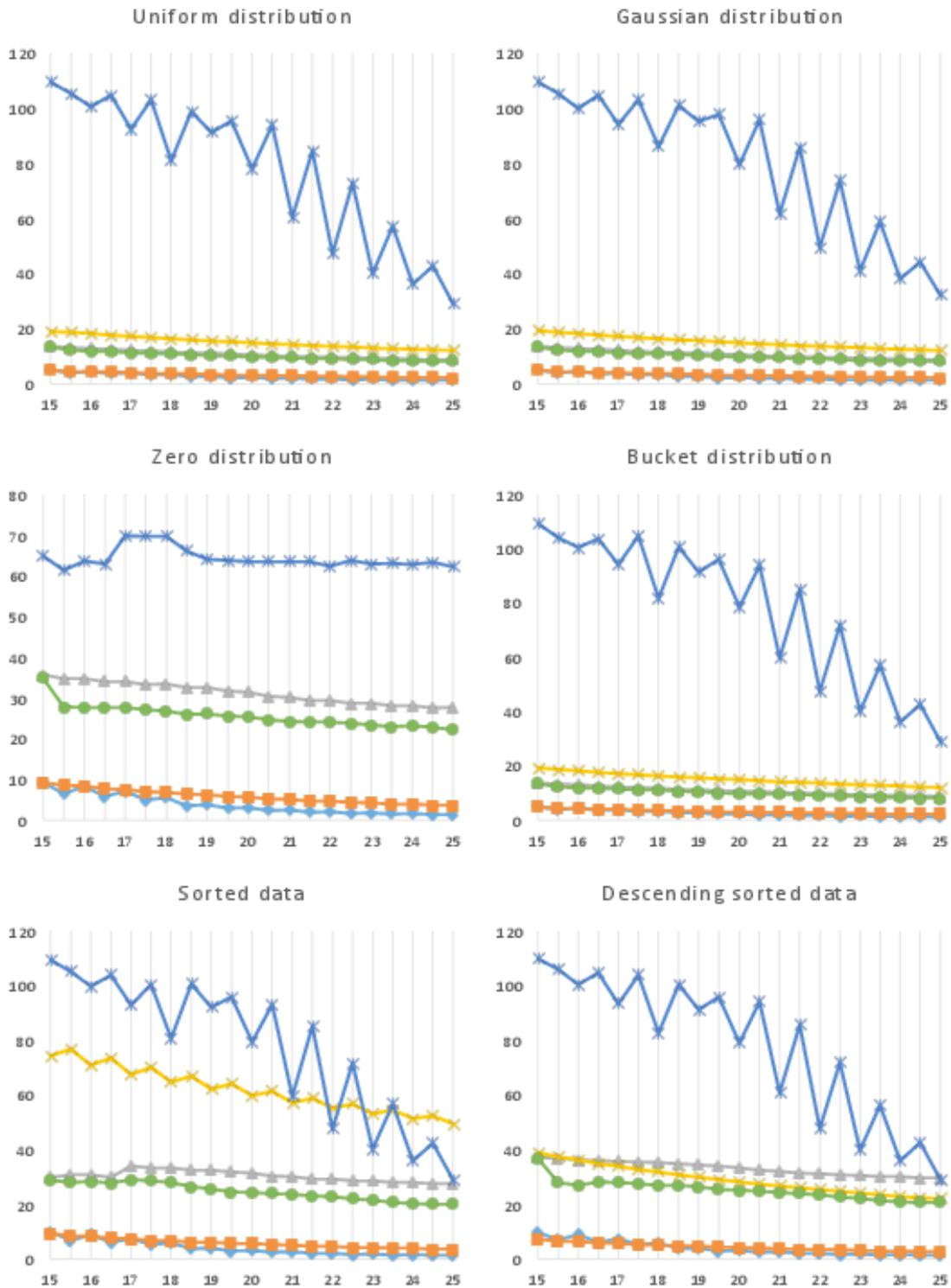

## Sequential, 32 bit, (key, value) pairs
The sort rate of sequential algorithms when sorting sequences of 32-bit (key, value) pairs.
X-axis: Binary logarithm of the sequence length, Y-axis: sort rate in M/s.

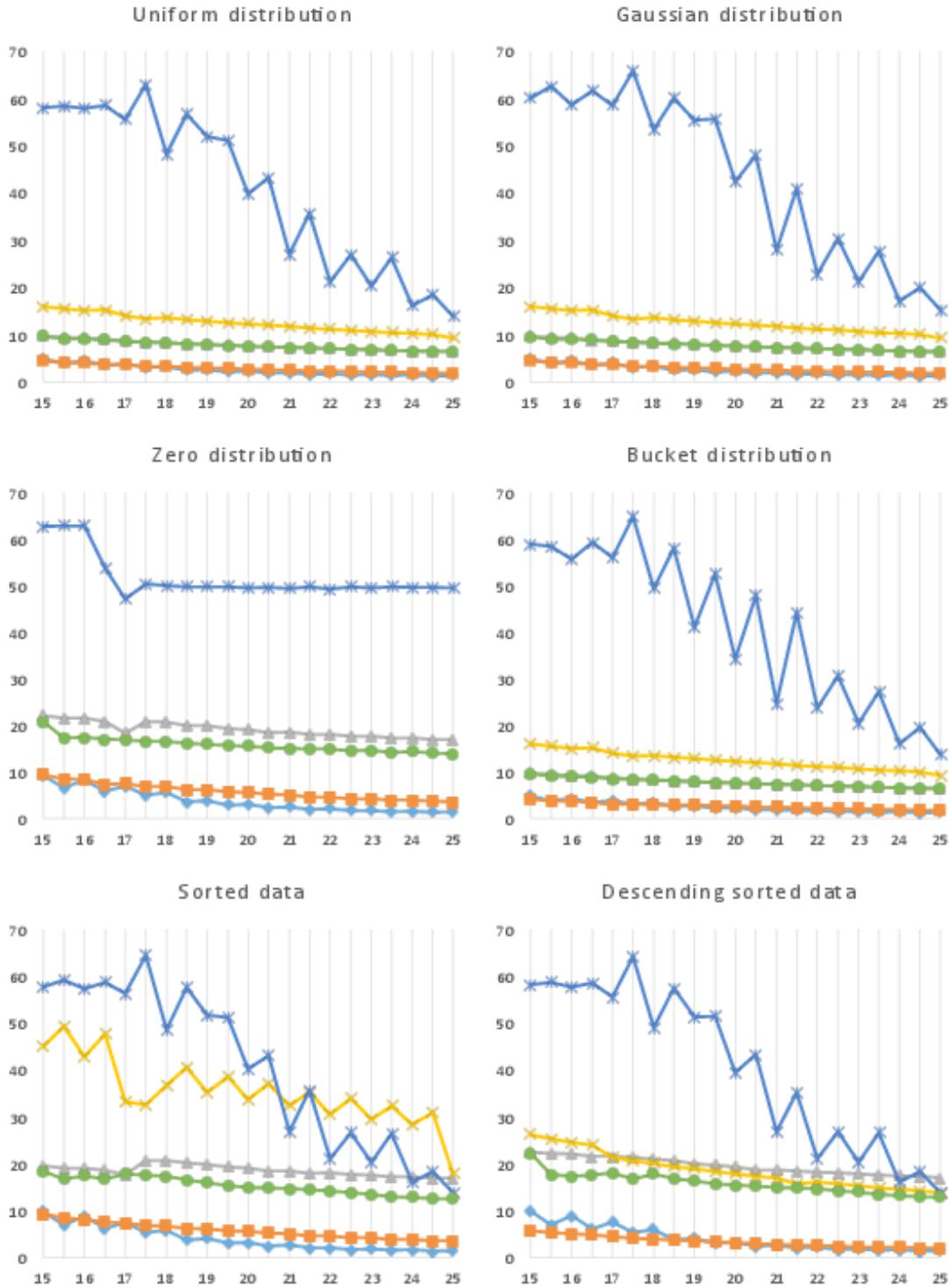

## Sequential, 64 bit, keys only
The sort rate of sequential algorithms when sorting sequences of 64-bit keys.
X-axis: Binary logarithm of the sequence length, Y-axis: sort rate in M/s.

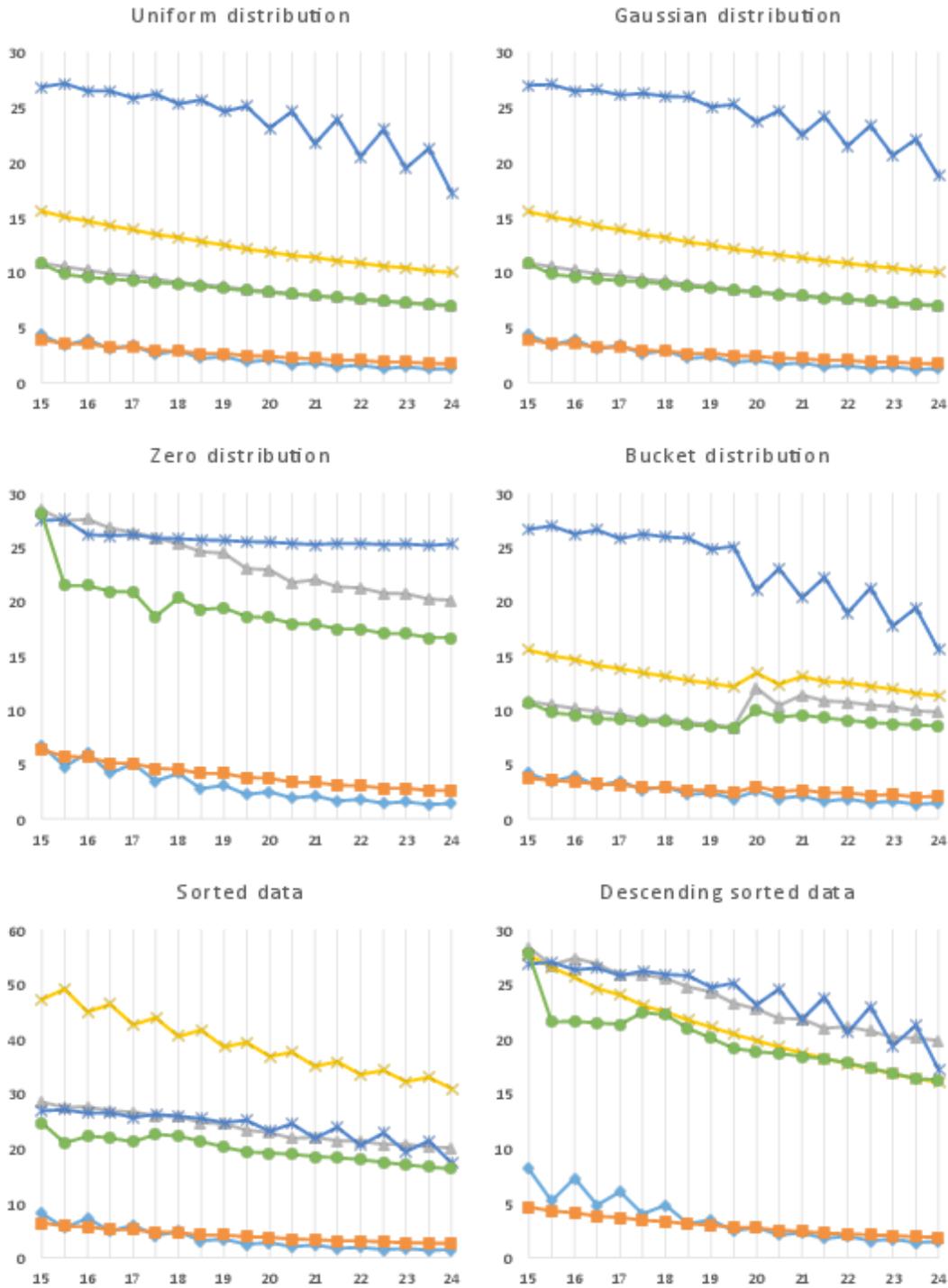

## Sequential, 64 bit, (key, value) pairs

The sort rate of sequential algorithms when sorting sequences of 64-bit (key, value) pairs.
X-axis: Binary logarithm of the sequence length, Y-axis: sort rate in M/s.

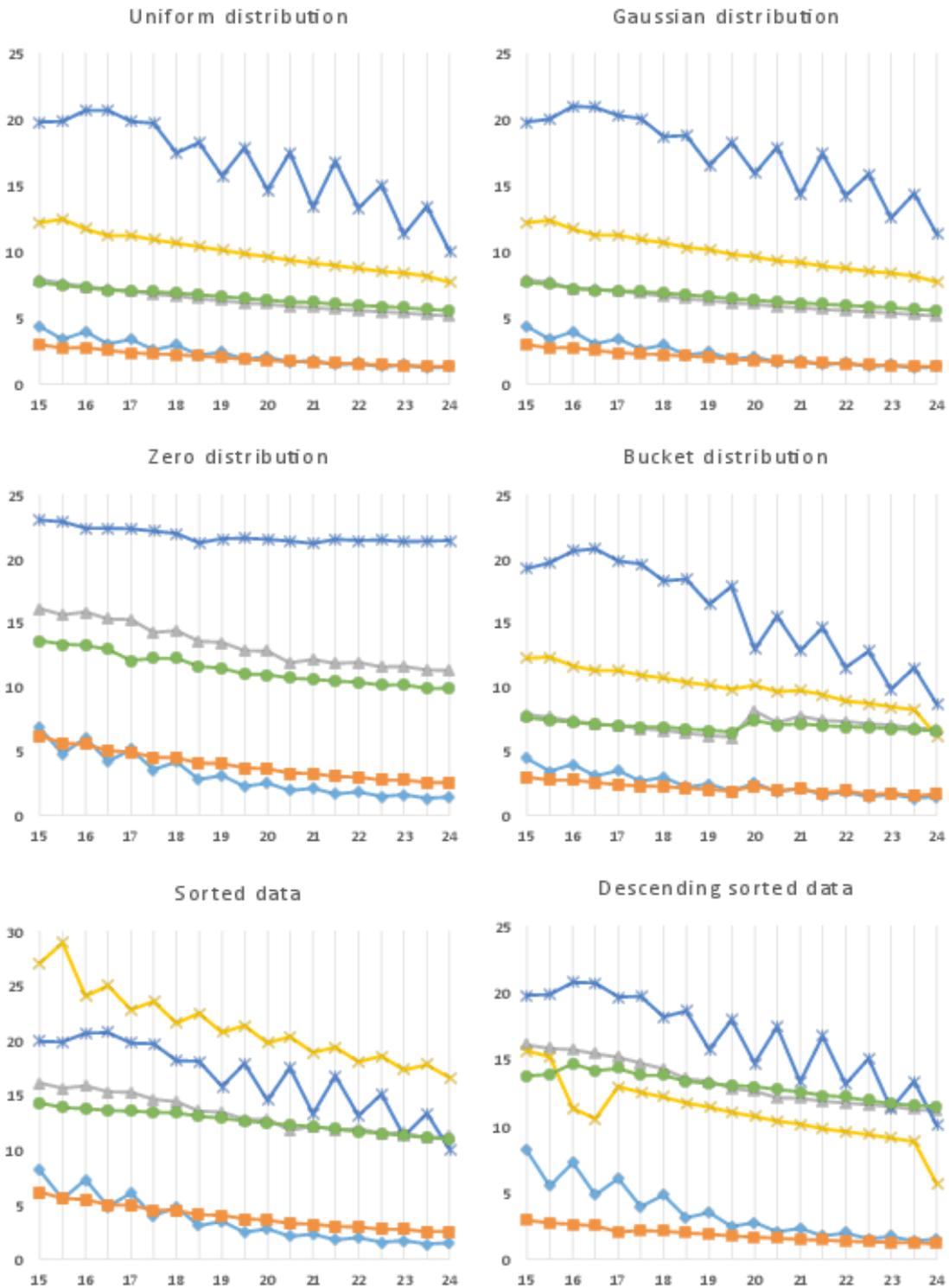